\documentstyle[11pt,newpasp,twoside,epsf]{article}
\def\edcomment#1{\iffalse\marginpar{\raggedright\sl#1\/}\else\relax\fi}
\reversemarginpar

\begin{document}
\title{Probing the Redshift Desert Using the Gemini Deep Deep Survey: observing galaxy
mass assembly at $z>1$}
 \author{Karl Glazebrook}
\affil{Department of Physics \& Astronomy, Johns Hopkins University,
Baltimore, MD 21218-2686, United States}
\author{\& The Gemini Deep Deep Survey team. (http://www.ociw.edu/lcirs/gdds.html)}
\affil{}

\begin{abstract}
The aim of the Gemini Deep Deep Survey is to push spectroscopic studies of complete galaxy samples (both
red and blue objects) significantly beyond $z=1$; this is the redshift  where the current 
Hubble sequence of ellipticals and spirals is already extant. In the Universe
at  $z=2$ the only currently spectroscopically confirmed galaxies are blue, star-forming and of fragmented morphology. 
Exploring this transition
means filling the `redshift desert' $1<z<2$ where there is a dearth of
spectroscopic measurements. 
To do this we need to secure redshifts of the oldest,
reddest galaxies (candidate ellipticals) beyond $z>1$ 
which has led us to carry
out the longest exposure redshift survey ever done: 
100 ksec spectroscopic MOS exposures with GMOS on Gemini North. We have developed an 
implementation of the CCD ``nod \& shuffle'' technique to ensure precise sky-subtraction in these 
ultra-deep exposures. At the halfway mark the GDDS 
now has $\sim 36$ galaxies in the redshift desert $1.2<z<2$ extending up to 
$z=1.97$ and $I<24.5$ with secure redshifts based on weak rest-frame UV absorption features complete 
for both red, old objects and young, blue objects. The peak epoch of galaxy assembly
 is now being probed by direct spectroscopic investigation for the first time. On behalf of the 
 GDDS team  I present our first results on the properties of galaxies in the `redshift desert'.
\end{abstract}

\section{Introduction: the `redshift desert'}

The `redshift desert' is the well known gap between $z=1$ and $z=2$ where there is a dearth of spectroscopic
redshifts. This is illustrated in Figure~1 which shows a color-redshift plot of some  of the deepest Keck spectroscopic
samples now available. Generally galaxies fall in too two groups: pure magnitude selected samples which now reach
as faint as $R_{AB}<24$ and redshifts $z<1.2$ and at higher redshifts the `Lyman Break' galaxies. These are color-selected
via their Lyman break and are easily found for $z>3$ (Steidel et al. 2003) and are being extended down
to $z=2$ (Erb et al. 2003). 

\begin{figure}[t]
\begin{center}
\plotfiddle{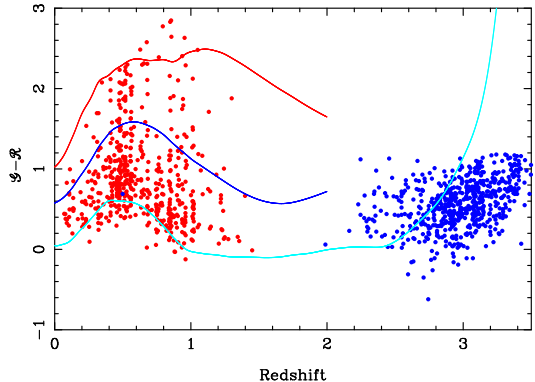}{1.5in}{-90}{30}{30}{-120}{170}
\caption{Current deep spectroscopic surveys. The  $z<1.2$ points are from Cohen et al. (2001)
and the $z>2$ points are Lyman break galaxies from Steidel et al. (1995). The tracks represent
the colors expected of non-evolving galaxy SEDs: E$/$S0, Sbc, Scd (red -- blue)}
\label{fig1}
\end{center}
\end{figure}

These samples are selected optically and as a result the galaxies selected at high-redshift have turned out to be those with the strongest ultraviolet (UV) emission because for $z>1$ the observed optical probes the rest-frame UV. Of note are the Lyman break galaxies which all have spectra characteristic of starburst galaxies: 
blue UV continua, ISM absorption lines and evidence of outflows. 

Other high-redshift populations, which emit less strongly in the UV, have been found. Examples include
the sub-mm selected SCUBA sources (Chapman et al. 2003) and the class of objects known as 
`Extremely Red Objects' selected via their strong near-infrared (NIR) emission and red optical-NIR
colors (e.g. Smail et al. 2002). While some spectra of such objects have been achieved there has been a 
lack of success in securing large samples of both red and blue galaxies above $z=1$.

The reason for this is primarily that the K-correction above $z=1$ has such a strong effect. Most galaxies
apart from strong starbursts are significantly less luminous in the near-UV than at optical wavelengths,
this is especially true of galaxies with SEDs characteristic of old stellar populations. Furthermore there
is a lack of strong emission lines, even in starforming galaxies, between [OII] (3727\AA) and 
Lyman-$\alpha$ (1216\AA). Thus spectroscopy must reach sufficient signal:noise to robustly
detect the continuum and hence achieve absorption line redshifts.

\section{Galaxy Spectroscopy with Nod \& Shuffle}

Our goal with the `Gemini Deep Deep Survey' (GDDS) was to achieve a 
complete NIR magnitude limited sample
of both red and blue galaxies pushing well above $z=1$ with the aim of reaching $z=2$. In order
to do this we need to be able to secure absorption line spectra of the reddest galaxies
---  candidate elliptical galaxies with redshifts $z\ga 1.5$ which are the hardest
to get redshifts of as all the flux comes out in the red and there are no emission lines. 
Therefore one needs efficient red spectroscopy. In principle the best option would
be near-IR 1--2\micron\ spectroscopy but too date there are no known examples of $z>1$ 
normal galaxies with absorption line NIR redshifts. This is because the sky is extremely bright
in the NIR and because the detectors are very noisy. Moreover the field-of-view of the instrumentation
is limited by the expense of NIR detector technology.

For the GDDS the compromise approach we adopted was to push on the red-end of the optical
passband (5000-10\,000\AA) using CCD technology. The readnoise and sky brightness are
reduced with respect to the near-IR, 
but so are the fluxes of the target objects. A simple estimate shows that a $L^*/3$ $z=1.5$
elliptical could be as faint as $I=24.5$ requiring a $\sim 100$ ksec spectroscopic exposure to
reach $S/N=5$ in the continuum at low resolution. Such long exposures are subject to many
possible systematic effects making sky removal difficult --- objects are $\la1\%$ of the sky. However
if such long exposures were possible then the optical approach would have a grant advantage:
because CCDs are relatively inexpensive instruments typically offer much wider fields-of-view and
hence greater multiplex advantage.

We adopted an approach known as `nod \& shuffle' (Glazebrook \& Hawthorn 2001) to facilitate
long exposures with accurate sky-subtraction in the red. (Figure~2). Briefly part of the CCD acts
as a storage area and  the telescope is nodded between an `object' and a `sky position'. (In
our case these were two neighboring positions along a slit). The spectrum at the sky position
is stored in the storage area by `charge shuffling' which introduces no additional noise provided
a CCD has good charge transfer efficiency.

\begin{figure}
\begin{center}
\plotone{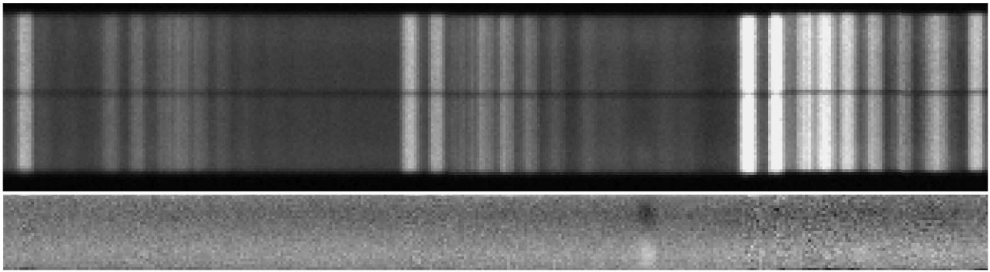}
\caption{\small Illustration of nod \& shuffle. The top panel shows a raw nod \& shuffle GDDS slit spectrum which consists
of two shuffled pairs. The  upper of the pair has the object at the top of the 2 arcsec slit (`A position'). 
The shuffled image just
below it has the object at the bottom of the slit (`B position'). Both images are observed quasi-simultaneously
through the same slits and CCD pixels, the lower image is shuffled down for storage. Bottom panel: the result of
simple image arithmetic A$-$B reveals a galaxy, with a strong emission line, visible continuum and Poisson
limited sky-subtraction. }
\label{fig:nodshuf}
\end{center}
\end{figure}

The nodding \& shuffling is typically done many times throughout the course of a long
exposure before reading out. Sky subtraction simply consists of subtracting the image from itself
after being shifted by the shuffle distance.

For the GDDS we cooperated with the Gemini Observatory to implement nod \& shuffle on
the optical GMOS spectrograph (Crampton et al. 2000). This is notable as the first implementation
of nod \& shuffle on an 8m telescope. Commissioning tests in July 2002 indicated that sky-subtraction
performance better than 0.1\% was being achieved with a 60s object$+$60s sky nod \& shuffle
cycle. For more details about nod \& shuffle with GMOS and the modes used with the
GDDS see Abraham et al. (2003).

\section{The Gemini Deep Deep Survey}

The scientific design of the GDDS required spectroscopy of a near-IR selected sample to $z=2$. 
Targets were selected from the Las Campanas Infrared Survey which has deep optical and
NIR data over many square degrees and has good photometric redshift estimates (Chen et al. 2003).
The advantages of near-IR selection are well known (Glazebrook et al. 1994). If one selects in the
$K$-band then the K-corrections are SED independent to roughly $z=2$. Moreover the light
is dominated by the  old stellar population, selecting in $K$ is a very good proxy for selecting by
stellar mass. Thus a near-IR selected sample can select massive galaxies at high-redshift
if they exist even if they are not actively star-forming and have weak UV. (The K20 survey
has similar motivations and goals but is less deep; see Cimatti et al., 2002.)
The survey design consisted of:

\begin{enumerate}
\item A primary $K<20.8$ selected sample. The photo-z distribution of the LCIRS indicated we
needed to go this deep in order to penetrate to $z=2$. For the reddest galaxies at this limit 
$I=24.5$.
\item A fill-in sample of low-priority targets, anything with $I<24.5$ where there was space on the mask.
\item Photometric redshifts were used to weed out objects with $z<0.8$. This has to be done
or else the low redshift foreground dominates. At low redshifts photo-z's work
the best because of the vastly better calibration, so this robustly leaves us with a $0.8<z<2$ sample.
\item Four fields were picked which lay within $5'\times 5'$ GMOS tiles, typically 70 slits could be allocated 
within each field. With nod \& shuffle the slits only needed to be 2 arcsec long.
\item Each field was observed for $\simeq 100$ ksec total using a 60s$+$60s nod \& shuffle cycle, reading out
every 1800s. Low resolution ($R\sim 500$) 5000--10,000\AA\ GMOS spectra were taken. 
\end{enumerate}

All four fields were observed on Gemini-North between August 2002 and July 2003 and have been
completely reduced. Redshifts and spectral classes were assigned by eye (by committee!) --- in particular
for the early-type galaxies by comparing with low-redshift templates. The final catalogs are currently being
prepared for publication.

Some example spectra are shown in Figure~3, absorption line redshifts are obtained for
both blue star-forming galaxies (ISM absorptions in FeII and MgII) and red early-type
spectra (from matching to low-redshift UV templates which exhibit the same broad
F features). 

\begin{figure}
\begin{center}
%\vspace{6in}
\plotone{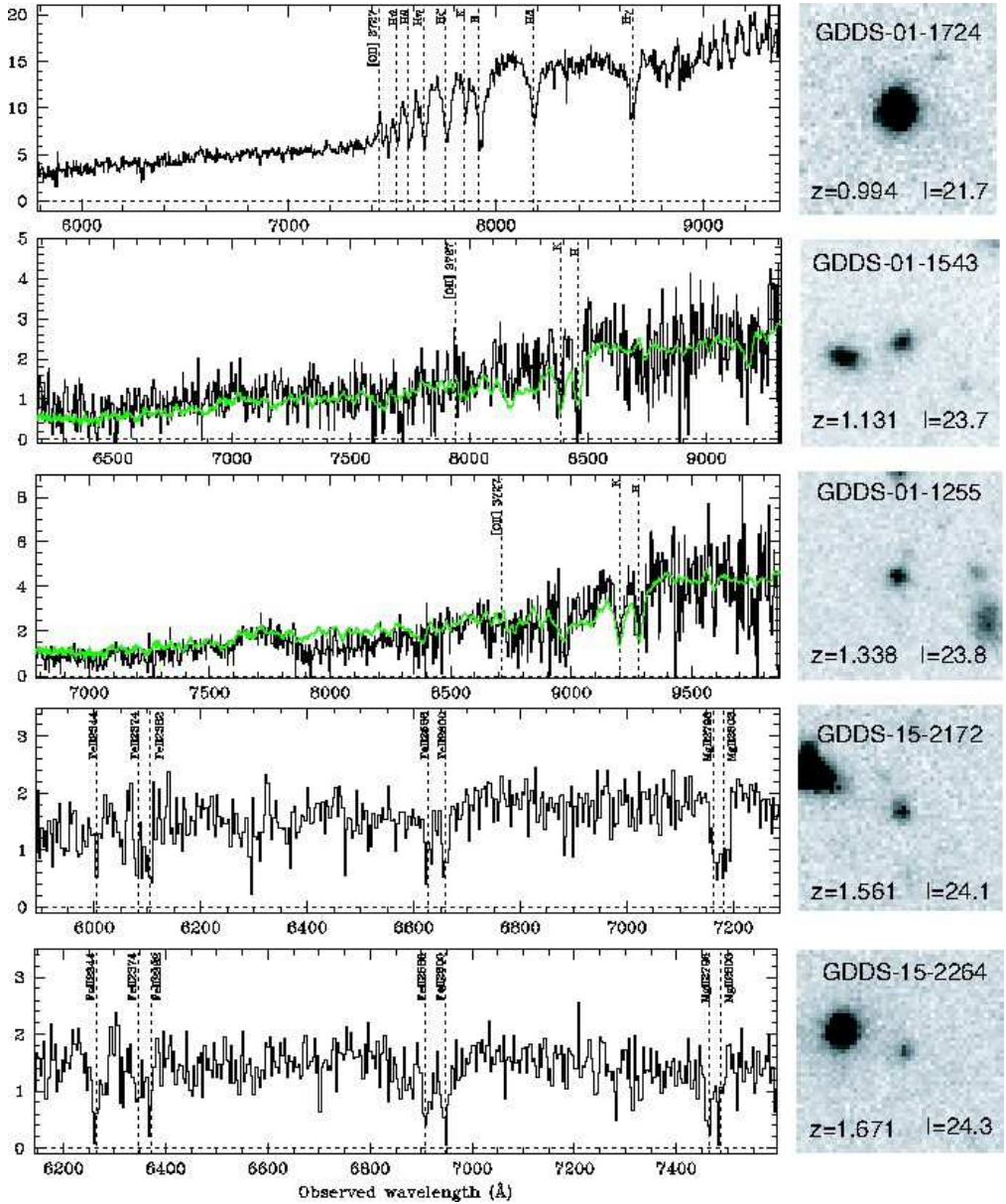}
\caption{Sample spectra from the GDDS (left) with corresponding ground based LCIRS images (right). 
Absorption line redshifts are obtained for both red and blue targets. The early type spectra
(01-1543 and 01-1255 here) are compared with a low-redshift templates. The broad features in
the spectra from F stars are well reproduced, which turns out to be the key to obtaining these
redshifts.}
\end{center}
\end{figure}

\section{GDDS early results}
\def\Msun{\hbox{M$_{\odot}$}}

Figure~4 shows the GDDS color-redshift plot from the first two fields for the $K$-selected
primary sample. The spectroscopic completeness is
77\%. There are 36 galaxies with spectroscopically confirmed galaxies with secure redshifts 
$1.2<z<2$.  It can be seen that we are picking up a complete range of SED types to $z\simeq 1.7$ and identifying
objects much redder than Lyman break galaxies at comparable redshifts.

\begin{figure}[t]
\begin{center}
\plotfiddle{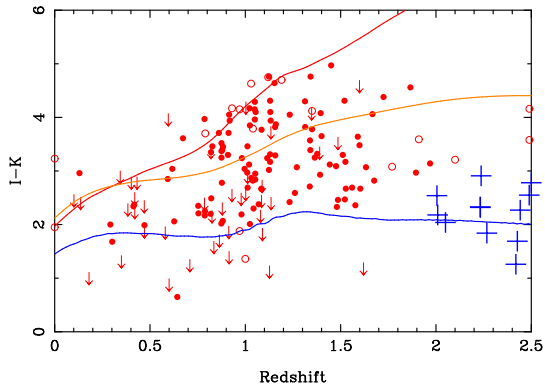}{1.5in}{-90}{30}{30}{-120}{170}
\caption{Color-redshift relation of GDDS galaxies in the first two fields (solid circies: confirmed spectroscopic
redshifts, open circles: unidentified spectra plotted at photometric redshifts)  compared with Lyman
Break galaxies (crosses, this data
supplied by Dickinson et al., private communication). Again the model tracks are non-evolving
E/S0. Sbc and starburst SEDs.}
\label{fig1}
\end{center}
\end{figure}

It is of course natural to ask `how massive are these galaxies' ? The typical Lyman Break objects
have stellar masses  $<10^{11}$\Msun\ (Papovich et al. 2001). We might expect that our objects
being much brighter in the $K$-band will be considerably more massive.

To answer this question we have fit SEDs to our multicolor data to determine $K$-band mass-to-light
values for the stellar population and hence derive stellar masses. Our procedure will be described in
detail elsewhere; however we note that to first order the $K$-band $M/L$ values vary much less than
they do in the optical (typically only a factor of two) and are far less 
sensitive to extinction and metallicity
uncertainties. We do use throughout a full calculation fitting over a grid of 95,864 model galaxy
SEDs computed using the PEGASE.2 code (Fioc \& Rocca-Volmerange 1997). These models include a full range of
dust extinction, metallicity, multiple bursts and different long-term star-formation histories.

\begin{figure}[t]
\begin{center}
\plotfiddle{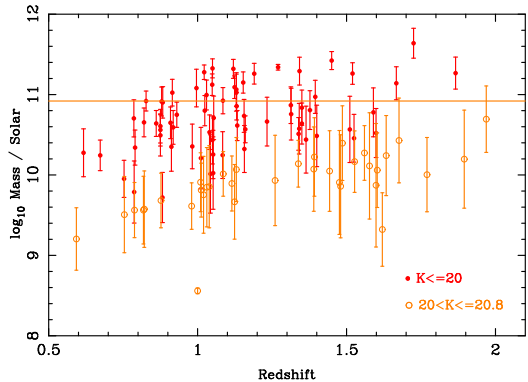}{1.5in}{-90}{30}{30}{-120}{170}
\caption{Mass(stellar)-redshift relation of GDDS galaxies in
the first two fields with different observed $K$-magnitude
ranges plotted as indicated. The solid horizontal line is the position of $M^*$ at $z=0$ from
Bell et al. (2003) }
\label{fig1}
\end{center}
\end{figure}

Our results are shown in Figure 5. This shows the masses of our galaxies as a function of our
$K$-band limit. It can be seen that for $K<20.8$ we reach a mass limit of  $>10^{10}$\Msun\
at $z=2$, and that a considerable number of galaxies have masses $> 10^{11}$\Msun, 
greatly in excess of the LBG population. This is natural given the galaxies are red and
$K$-luminous. Many of these are spectroscopically confirmed to have old stellar populations.
Such an abundance of massive galaxies may present a considerable problem for $\Lambda$CDM
models of hierarchical galaxy formation which typically predict a factor of $\sim 30$ drop in 
the number of $10^{11}$\Msun\ galaxies between $z=1$ and $z=2$. (Baugh et al. 2003).

We are currently working on finalizing the space density of these massive galaxies at $z>1.5$,
this work will be published as Glazebrook et al. (2004).
The beauty of the GDDS is that we have spectra as well. We can assess whether a 
galaxy is spectroscopically old or a reddened starburst. We can also determine
abundances from the spectra (see Savaglio et al. 2004, in press).

\section{Conclusions}

The Gemini Deep Deep Survey has succeeded in obtaining spectroscopic redshifts
of galaxies in the former redshift desert.  We have closed the $1<z<2$ gap, but 
more importantly we are highly complete for both red and blue
galaxies. A catalog paper (Abraham et al. 2004) is in preparation. Our results to date are:

\begin{enumerate}
\item An abundance of massive galaxies at $z\ga 1.5$. (Glazebrook et al. 2004, in prep.) These
may be problematic for $\Lambda$CDM models of heircharchical growth of massive halos.
\item Many of the massive red galaxies at $z>1$ have spectra of pure old stellar populations (McCarthy et al. 2003, in
prep.)
\item Massive starburst galaxies at $1.3<z<2$ have high metal columns and are close to solar in
abundances (Savaglio et al. 2004 in press).
\end{enumerate}

At the time of this meeting we had just received the Gemini data for the final two
of our four fields. The complete
sample will be more than double the size of the one presented here.
To come we have been awarded Cycle 12 time to image our fields with the Advanced Camera
for Surveys on the Hubble Space Telescope. Thus we hope to connect our results to
physical galaxy morphology. It is clear we are now living in an age where the formation of galaxies
and the Hubble Sequence can be observed directly.

\section{Acknowledgments}

This work is based on observations obtained at the Gemini Observatory, which is operated by the
Association of Universities for Research in Astronomy, Inc., under a cooperative agreement
with the NSF on behalf of the Gemini partnership: the National Science Foundation (United
States), the Particle Physics and Astronomy Research Council (United Kingdom), the
National Research Council (Canada), CONICYT (Chile), the Australian Research Council
(Australia), CNPq (Brazil) and CONICET (Argentina). Karl Glazebrook acknowledges
genereous funding from the David \& Lucile Packard Foundation.

\end{document}